\documentclass[12pt,preprint]{aastex}

\newcommand{\myemail}{t.schmidt@chem.usyd.edu.au}

\shorttitle{hexabenzocoronene} \shortauthors{Kokkin et al.}

\begin{document}

\title{The optical spectrum of a large isolated polycyclic aromatic
hydrocarbon: hexa-\textit{peri}-hexabenzocoronene, C$_{42}$H$_{18}$}

\author{Damian L. Kokkin$^a$, Tyler P. Troy$^a$, Masakazu Nakajima$^a$, Klaas
Nauta$^a$, Thomas D. Varberg$^b$, Gregory F. Metha$^c$, Nigel T.
Lucas$^a$ and Timothy W. Schmidt\altaffilmark{a,1}}

\affil{$a$ School of Chemistry, The University of Sydney, NSW 2006,
Australia\\$b$ Department of Chemistry, Macalester College, St.
Paul, MN 55105 USA\\$c$ School of Chemistry \& Physics, The
University of Adelaide, SA 5005 Australia\\}

\altaffiltext{1}{Currently visiting researcher (CNRS) at Laboratoire
Photophysique Mol\'eculaire, Universit\'e Paris-Sud XI, Orsay 91405,
France. Author to whom correspondence should be addressed. \myemail}

\begin{abstract}
The first optical spectrum of an isolated polycyclic aromatic
hydrocarbon large enough to survive the photophysical conditions of
the interstellar medium is reported. Vibronic bands of the first
electronic transition of the all benzenoid polycyclic aromatic
hydrocarbon hexa-\textit{peri}-hexabenzocoronene were observed in
the $4080-4530\,${\AA} range by resonant 2-color 2-photon ionization
spectroscopy. The strongest feature at 4264\,\AA\ is estimated to
have an oscillator strength of $f=1.4\times10^{-3}$, placing an
upper limit on the interstellar abundance of this polycyclic
aromatic hydrocarbon at $4\times10^{12}$\,cm$^{-2}$, accounting for
a maximum of $\sim0.02$\% of interstellar carbon. This study opens
up the possibility to rigorously test neutral polycyclic aromatic
hydrocarbons as carriers of the diffuse interstellar bands in the
near future.
\end{abstract}


\section{Introduction}

Organic material such as polycyclic aromatic hydrocarbons (PAHs) are
held responsible for infrared emission features in carbon rich
objects (AIBs) and has been suggested to account for as much as 20\%
of interstellar carbon \citep{Leger1984,Allamandola85,Snow1995}. In
addition, PAHs are considered leading candidates as carriers of the
diffuse interstellar bands \citep{Leger1985,Crawford85,Salama1999},
a series of diffuse absorption features superimposed on the
extinction curve of the interstellar medium (ISM). The identities of
the carriers of these bands remain the longest unsolved problem of
laboratory astrophysics \citep{Herbig1995,Sarre2006}. A scenario
linking the AIBs to the DIBs is that amorphous aromatic material
formed in carbon rich stellar outflows is further processed into
PAHs and eventually the DIB carriers by the interstellar
photophysical environment \citep{Goto2003, Sloan2007, pino08}.

Despite the extensive literature reporting on PAHs in the ISM, not a
single PAH species has ever been observed astronomically. This may
be partly due to a paucity of laboratory data with which to compare
unidentified features in astronomical spectra, or to guide
astronomical searches. Obtaining optical spectra of isolated
molecules at low temperature (as in diffuse clouds), \emph{in
vacuo}, remains a challenge of modern laboratory astrophysics
\citep{Sharp2005}.

Modelling of generic PAHs in the diffuse interstellar medium aims to
guide the laboratory search for the carriers of the DIBs.
\citet{Leger1985} found that species must contain no fewer than
fifteen atoms if they are to avoid photo-thermolysis and
\citet{Lepage2003} concluded that PAHs containing more than about 30
carbon atoms would be largely stable. Furthermore, it is considered
that a large proportion of interstellar PAH material would exist in
its ionized form(s) \citep{Crawford85}. As such, the challenge to
laboratory-based scientists is to place large PAHs into the vacuum,
cool them down, perhaps ionize them and obtain their spectra. Prior
to the present work, the largest neutral PAH
 studied spectroscopically in such a way was ovalene
(C$_{32}$H$_{14}$), the size of which is borderline for surviving
the ISM intact \citep{Amirav1980}.

\begin{figure}[h]
\begin{center}
\includegraphics[width = 16cm]{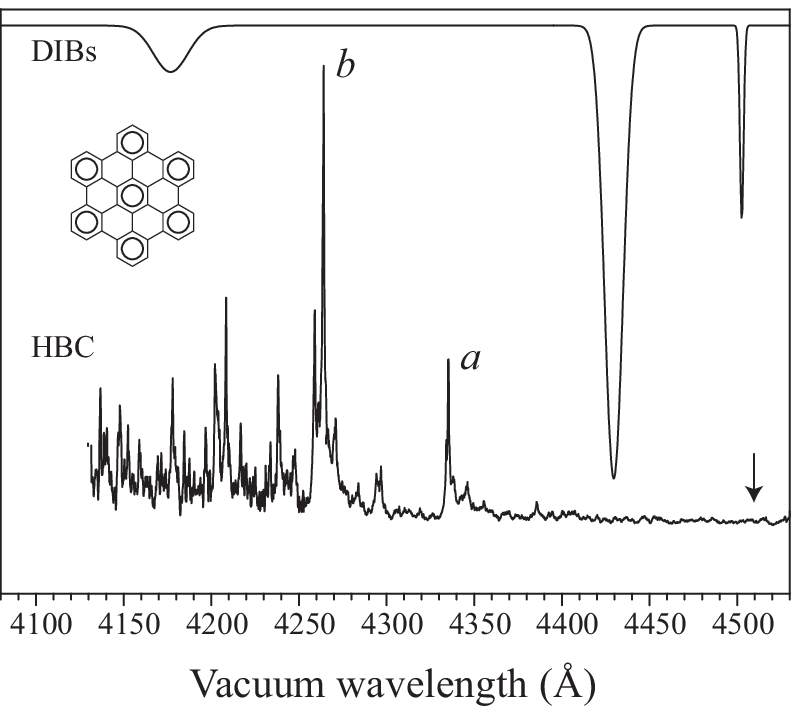}
\caption{\label{fig} The excitation spectrum of
hexa-\textit{peri}-hexabenzocoronene (bottom) and a reconstruction
of the diffuse interstellar band spectrum (top) from
\citet{Jenniskens1994}. The arrow denotes an estimate of the the
position of the forbidden origin band, based on the solution phase
spectra of \citet{Hendel1986}.}
\end{center}
\end{figure}

An unguided laboratory search is daunting, however, as for systems
containing $4-10$ fused aromatic rings there are over 20000 possible
PAH structures \citep{Dias2004}. In order to guide the laboratory
search for DIB carriers, some selection mechanism is required, based
on spectral intensity, or structure. Recently we suggested that
all-benzenoid PAHs (ABPAHs) could represent such a selection
mechanism \citep{Troy2006}. These ABPAHs are formed from fused
benzene rings which may be drawn as being separated by single bonds.
They are characterized by higher ionization potentials and more
energetic electronic transitions than their non all-benzenoid
isomers. Of the more than 20000 possible structures containing
$4-10$ fused rings, only 17 are all-benzenoid in character. The
all-benzenoid structural motif thus provides a selection mechanism
for the laboratory and perhaps the ISM. Since the ABPAHs are
comparatively easy to prepare in the laboratory, it may also be that
they form preferentially in the ISM. We obtained the gas phase
spectrum of the smallest member of this family: triphenylene
(C$_{18}$H$_{12}$)\citep{Kokkin2007}. However, its small size means
that even its lowest energy transitions are in the ultraviolet, far
from the shortest wavelength DIBs. The ABPAHs present three optical
band systems named the $\alpha$, $p$ and $\beta$-bands in increasing
intensity and energy. We found that for the strong $\beta$-band
system to fall in the DIB region, it must contain no fewer than 84
carbon atoms \citep{Troy2006}. However,
Hexa-\emph{peri}-hexabenzocoronene (HBC) is known to exhibit the
weaker $\alpha$-bands in the region of the DIBs. HBC , pictured in
Fig. \ref{fig}, possesses 42 carbon atoms and is thus large enough
to survive the interstellar radiation field, as modelled by
\citet{Lepage2003}. Since its lowest energy transitions occur in the
visible region, near the strongest DIB (4429\,\AA\ in air) it was
suggested as a carrier of this interstellar band by
\citet{Hendel1986} who obtained spectra in a 1,2,4-trichlorobenzene
solution. Furthermore, its red phosphorescence suggested that it
could also be a carrier of the \emph{Red Rectangle} Bands, the
unidentified emission features superimposed on the extended red
emission of the protoplanetary nebula surrounding HD 44179
\citep{Cohen1975,Schmidt1980}. \citet{Hendel1986} also measured the
mass spectrum of HBC and found that while HBC$^{2+}$ and HBC$^{3+}$
were abundantly produced, there was very little H-loss, suggesting
it to be a very tough molecule of the sort required to survive harsh
interstellar radiation. Despite these suggestions, a gas phase
spectrum of HBC has never been reported. In this letter we present
the gas phase spectrum of jet-cooled, neutral
hexa-\emph{peri}-hexabenzocoronene (HBC, C$_{42}$H$_{18}$). From
these measurements an upper limit on the column density of HBC in
the diffuse interstellar medium (DISM) is obtained.

\section{Experimental Methods}

HBC was synthesized in our laboratory using the general procedure of
\citet{Fechtenkotter1999} for the synthesis of C$_6$-symmetric HBC
derivatives. We estimate the purity of our sample as $>97$\% based
on MALDI-TOF mass spectrometry.

The target species was seeded in argon by ablating the solid HBC
sample with laser pulses (3.5\,mJ/pulse, 532\,nm) produced by a
Nd:YAG laser system. The seeded carrier gas was then expanded into
the source chamber where cooling occurred in a supersonic expansion.
The free jet was collimated with a 2\,mm diameter skimmer and passed
into the extraction chamber.

The jet-cooled HBC molecules were cooperatively ionized by two laser
pulses. The wavelength of the first laser pulse was tuned to place
HBC molecules into an excited state from where they could be ionized
by the second, fixed wavelength pulse. Tunable radiation was
provided by a 308 nm pumped dye laser operating with Exalite 417 and
Coumarin 440 laser dyes to scan the region of interest, 408-455 nm
($2.73-3.04$\,eV). The laser was weakly focused into the apparatus.
The ionization pulse was provided by the doubled output of an
optical parametric oscillator, pumped by a Nd:YAG laser system,
producing 213\,nm (5.82\,eV) which is able to ionize the excited
state HBC: The reported ionization potential of HBC is 6.86\,eV
\citep{Hendel1986}. This beam was aligned so as to counter-propagate
the tunable radiation. The nozzle and laser were pulsed at 10\,Hz.

The positive ions were extracted vertically and perpendicular to the
laser and molecular beam into the time-of-flight tube. The
accelerating voltage was 2050\,V, applied to the bottom extraction
grid. Ions were detected with a tandem microchannel plate.

The signal from the microchannel plate was buffered by a digital
oscilloscope and processed on a desktop computer running in-house
software. The synchronization and triggering of all the equipment
was carried out using a multiple output programmable delay
generator.

\section{Results and Discussion}
The resonant 2-color 2-photon ionization spectrum of HBC is plotted
in Fig. \ref{fig}. While an attempt was made to normalize the
spectrum for laser power, relative intensities become increasingly
unreliable towards 4100\,\AA. The spectrum reported here is for the
522 amu ($^{12}$C$_{42}\,^{1}$H$_{18}$) mass only, and as such the
spectrum represents only the most abundant isotopomer (64\%). A
detailed analysis of the spectrum, including the other isotopomers,
will be published elsewhere.

The main features are positioned at vacuum wavelengths of 4335.2 and
4264.1\,\AA\ (FWHM 1.5\,\AA). These are labeled $a$ and $b$ in Fig.
\ref{fig}. The band width is related to the rotational temperature
which is typically 30\,K in our experiments. We attribute the
cluster of smaller peaks immediately to longer wavelength of these
main features to `hot' bands, which are transitions arising from
excited vibrational levels of the electronic ground state that
remain populated by incomplete cooling in the supersonic expansion.
To higher energy than these main features lie transitions involving
quanta of totally symmetric vibrations built onto bands $a$ and $b$.

Preliminary quantum mechanical considerations reveal, that when
viewed on the central ring, the highest occupied molecular orbitals
and the lowest unoccupied molecular orbitals are the same as those
of benzene, which has the same symmetry ($D_{6h}$ point group). As
such, the symmetries of the lowest transitions of HBC are the same
as benzene. The lowest energy transition is thus of
$B_{2u}\leftarrow A_{1g}$ electronic character, induced by
vibrational modes of $e_{2g}$ symmetry. An estimate of the position
of the forbidden origin band is indicated by an arrow in Fig.
\ref{fig}. A detailed quantum mechanical treatment is beyond the
scope of the present article and will be presented elsewhere. We
assign the bands $a$ and $b$ to false origins arising from single
quanta of vibrational modes of $e_{2g}$ symmetry. An estimate of the
oscillator strength of the strongest band, $b$, may be made by
integrating over the molar extinction coefficient,
$\epsilon(\lambda)$, of the strongest visible region band in the
solution phase spectra of \citet{Hendel1986}, at 4450\,\AA,
\begin{equation}
f
=\frac{4.28\times10^{-11}}{\lambda^2}\int\epsilon(\lambda)d\lambda,\nonumber
\end{equation}
with $\lambda$ in m and $\epsilon$ in L\,mol$^{-1}$\,cm$^{-1}$.  As
such, the oscillator strength of $b$ is estimated to be
$f=1.4\times10^{-3}$.

Within the region of the spectrum plotted in Fig. \ref{fig} there
are several DIBs. These are plotted as negative going peaks above
the experimental spectrum with vacuum wavelengths of 4177.6, 4430.1
and 4503.0\AA. It is immediately clear that the experimental
spectrum does not match any hitherto reported DIB. It is also clear
that the 4177.6 and 4430.1\,\AA\ DIBs in particular are very broad
compared to the observed band and it is therefore unlikely that
these interstellar bands arise from transitions to the first excited
state of medium-sized PAHs such as HBC. Transitions to higher
electronic states will exhibit broader bands and stronger spectra.

The non-observation of HBC in the DIB spectrum allows us to estimate
an upper limit of the column density of HBC in the DISM using
\begin{equation}
N_{max} = 1.13\times10^{20}\frac{W_{max}}{\lambda^2 f}
\mathrm{\,cm}^{-2},\nonumber
\end{equation}
with $\lambda$ and $W_{max}$ in \AA. 
From our estimated oscillator strength, assuming a detection limit
of 1\,m{\AA} equivalent width, an upper limit on the column density
of HBC may be placed at $N_{max}=4\times10^{12}$\,cm$^{-2}$. Taking
the total carbon column density of $N_C=1\times10^{18}$ in a typical
DIB sightline, the HBC neutral represents a maximum of
$2\times10^{-4}$ as a fraction of interstellar carbon.

However, due to the exceptional properties of HBC an astronomical
search for the molecule is warranted. The next step in the
laboratory is to obtain the spectrum of HBC$^{n+}$, which is a more
difficult task in the gas phase. Nevertheless, the present work
makes it possible to produce cold cations in the gas phase by
threshold ionization through the 4261\,\AA\ band.

\section{Conclusions}

We have succeeded in obtaining the excitation spectrum of
hexa-\emph{peri}-hexabenzocoronene (C$_{42}$H$_{18}$) as a
jet-cooled, isolated molecule in the vacuum. Its strongest
transition in the region studied was found to lie at 4261.1\,\AA,
which does not correspond to any known diffuse interstellar band. An
estimate of the oscillator strength of this band was made at
$f=1.4\times10^{-3}$, which afforded an upper limit on the
interstellar abundance of HBC of $4\times10^{12}$\,cm$^{-2}$, or
0.02\% of interstellar carbon. However, the same technique can be
used to obtain the spectra of still larger and less symmetric
systems which may exhibit much stronger transitions. The collection
and analysis of the spectra of these all benzenoid PAH species
builds up the database of gas phase spectra required to constrain
models of PAH abundances in the ISM in concert with detailed
observation. In particular, an exploration of larger PAHs with
strong ($f\sim1$) transitions will provide a rigorous test of
neutral PAHs as carriers of the DIBs, thus opening the door to
solving the longest standing mystery in astronomical spectrocopy, or
putting to bed a long-favored class of candidate.

There remains an impetus to collect spectra of cold PAH cations. The
present work establishes the ability to place PAHs of 42 carbons in
the vacuum by laser desorption and supersonic expansion. The next
step is to produce the cold cations by threshold ionization and
obtain their spectra by multiphoton dissociation \citep{Pino2007}.



\acknowledgments

We are grateful to Thomas Pino for helpful discussions while
preparing this manuscript. DLK and TPT thank the University of
Sydney for a University Postgraduate Award. KN and NTL acknowledge
the Australian Research Council for the award of an Australian
Research Fellowship and an Australian Postdoctoral Fellowship.
Acknowledgment is made by TDV to the Donors of the American Chemical
Society Petroleum Research Fund for partial support of this
research. This research was supported under the Australian Research
Council's Discovery funding scheme (project numbers DP0665824,
DP0665831, DP0346380, DP0666236 and DP0556336) and the Australian
Research Council's Linkage Infrastructure Equipment and Facilities
funding scheme (project number LE0560658).

\clearpage

\end{document}